\documentclass[prb,superscriptaddress,reprint,amssymb,aps,floatfix,showkeys]{revtex4-1}
\usepackage{amsmath}%
\usepackage{amsfonts}%
\usepackage{amssymb}%
\usepackage{graphicx}
\usepackage{color}
\usepackage{tabularx}
\usepackage{braket}

\begin{document}
\title{Effects of Geometry on Near Quantum Ground State Behaviour of Phonon-Trapping Acoustic Cavities}

\author{Maxim Goryachev}
\email{maxim.goryachev@uwa.edu.au}
\affiliation{ARC Centre of Excellence for Engineered Quantum Systems, University of Western Australia, 35 Stirling Highway, Crawley WA 6009, Australia}

\author{Michael E. Tobar}
\email{michael.tobar@uwa.edu.au}
\affiliation{ARC Centre of Excellence for Engineered Quantum Systems, University of Western Australia, 35 Stirling Highway, Crawley WA 6009, Australia}

\begin{abstract}
This work presents some peculiarities of the near quantum ground state behaviour of curved (phonon trapping) Bulk Acoustic Wave (BAW) cavities when compared to a conventional mechanical resonator. The curved cavity system resolves the quandary of the conventional mechanical system where the Bose-Einstein distribution requires higher frequencies for lower quantum occupation factors contrary to the constraint of an inverse frequency dependence of the quantum fluctuations of displacement. We demonstrate how the non-trivial cavity geometry can lead to better phonon trapping, enhancing the variance of zero-point-fluctuations of displacement. This variance becomes independent of overtone number (or BAW resonance frequency) overcoming the constraint and allowing better observation of quantum effects in a mechanical system. The piezoelectric electro-mechanical coupling approach is qualitatively compared to the parametric optomechanical technique for the curved BAW cavities. In both cases the detectible quantity grows proportional to the square root of the overtone number, and thus the resonance frequency. Also, the phonon trapping improves with higher overtone numbers, which allows the electrode size to be reduced such that in the optimal case the parasitic capacitive impedance becomes independent of the overtone number, allowing effective coupling to very high frequency overtones.
\end{abstract}
%\date{\today}

%\pacs{07.64.+z, 07.10.Cm, 03.65, 43.35.+d}
% Keywords required only for MST, PB, PMB, PM, JOA, JOB? 
%\vspace{2pc}
%\noindent{\it Keywords}: Phonon trapping, Acoustic cavity, Quantum ground state, Cryogenic temperatures\\
% Uncomment for Submitted to journal title message
%\submitto{\NJP}

\maketitle

%\begin{multicols}{2}

\section*{Introduction}
 
Bulk Acoustic Wave (BAW) devices at cryogenic temperatures demonstrate great potential for many physical applications\cite{ScRep} exhibiting quality factors over a billion\cite{quartzPRL} at frequencies approaching $1$ GHz. In particular, they show great promise to operate as a mechanical system at the quantum limit with an extremely high-Q\cite{Aspelmeyer:2008qc,schwab}. These devices have the largest $Q\times f$ product among all the devices cooled to near their quantum ground state\cite{Kippen}. Nevertheless, a detailed description of BAW devices is not well known outside the acoustic device community. In particular, questions about mode masses and piezoelectric detection of the mechanical vibration remain outside the scope of recent experimental work\cite{apl1,Goryachev1,ScRep,quartzPRL}. This article has a purpose to answer these questions in terms of condensed matter physics, as well as to give an introduction to the detailed literature, which already exists. 

Being simultaneously a mechanical (acoustic) resonator and a phonon analogue of the Fabry-P\'{e}rot cavity, these devices demonstrate a number of interesting features that cannot be seen in standard mechanical resonators like membranes or cantilevers. In addition to this cavity-like nature, some additional peculiarities are due to the special phonon-trapping plate geometry used to drastically reduce the resonator coupling to the environment and thus to achieve the outstanding results in terms of quality factors. Hence it is important to demonstrate the role of the nontrivial geometry of the device, its impact on the Harmonic Oscillators (HO) corresponding to the different mode structures within the BAW cavity and to compare them to modes in a trivial (flat) plate case. The latter type of an acoustic cavity is represented, for example, by Film Bulk Acoustic Wave Resonators (FBAR) or High Overtone Bulk Acoustic Wave Resonators (HBAR). The former device has been recently cooled to the ground state\cite{fbar}, revealing a lack of coherence time, i.e. quality factor, for the full quantum state spectroscpopy. This work utilises a similar piezoelectrical detection approach, which is an alternative to standard optomechanics.  
 
As a starting point, we utilise the result of Stevens and Tiersten\cite{Tiersten1969,stevens:1811,Tiers1}. These results have been verified experimentally over the last few decades and resulted in the prosperity of piezoelectric BAW technology\cite{1537081,Goryachev:2013ly} as a building block for the time-keeping community. Although, in order to simplify the detailed and complicated calculations of the original authors, some additional approximations are made. In particular, an assumption of weakly anisotropic material is used throughout the work. Thus, all the presented calculations should be considered as rough estimates for such anisotropic materials such as quartz. Nevertheless, the demonstrated principles are fundamental for the curved BAW devices. 
{It should also be highlighted here that the main objective of this work is to analyse an acoustic system with the described 'Fabry-P\'{e}rot like' geometry, rather than to make numerical calculations for quartz BAW resonators. The analysis is not limited to this material or even to piezoelectric crystals in general. The main difference for the non-piezoelectric case is the impossibility of this type of detection as described in Subsection~\ref{piezo}.}

\section{General Description of Curved BAW Resonators}
\label{this}

\begin{figure}[htb!]
\centering
            \includegraphics[width=0.45\textwidth]{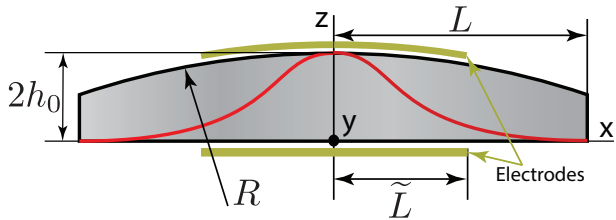}
%            \end{center}
    \caption{Side view of a curved BAW cavity. Red curve shows typical distribution of a mode the displacement along the cut in the case $m=0$ and $p=0$.}%
   \label{F0041FS}
\end{figure}

In this work, we consider a curved BAW piezoelectric plate device (Fig.~\ref{F0041FS}). Such a device confines the acoustical waves to the central region of the plate, which propagate along the thickness of the $z$-axis. Typically three types of so-called thickness acoustical waves could be excited. They are longitudinal (A-mode), fast shear (B) and slow shear (C) waves. Typically a crystal plate exhibits resonances at different frequencies corresponding to different modes and wave numbers. Summarising the present day knowledge of such systems, a resonance is characterised by an overtone number $n$ (showing how many half waves are in the plate thickness) and two other wave numbers $m$ and $p$ characterising the distribution of the energy maxima in the resonator $x$-$y$ plane. {In general, assigning the indices $m$ and $p$ assumes a regular separable solution and hence mode shape.}

 It is known that not all of these modes could be excited piezoelectrically. In other words, not all of them are coupled to the piezoelectrical environment. Such modes could be excited only mechanically, for example, by another mode through nonlinear coupling. Such hidden modes are those with even overtone (OT) number $n$, or odd energy distribution numbers $m$ or $p$. For this reason only modes with odd $n$ and even $m$ and $p$ are considered in this work. Normally, the operational overtone is the fundamental mode ($m=p=0$) with only one energy maximum, which is in the centre of the crystal. 

{It is also important to underline the main differences between room-temperature operation of BAW devices and their use at cryogenic temperatures\cite{ScRep}. Whereas in the former temperature range, devices are designed in order to achieve maximum middle and long term frequency stability, in the latter range we are mainly concerned with maximising the  $Q\times f$-product. 
The room temperature applications require such devices to operate with low order overtones of the shear modes, where maximum values of the $Q$-factors are achieved. In contrast, cryogenic operation provides an opportunity to operate at extremely high OT numbers of the longitudinal mode\cite{quartzPRL}. This difference is primarily due to different loss mechanisms at these temperatures. Firstly, room temperature operation means that acoustic losses are limited by the Akheiser mechanism\cite{Akheiser} that implies $Q\times f =  \mbox{const}$, while at cryogenic temperatures the Landau-Rumer\cite{landaurumer1} mechanism dominates with a $Q =  \mbox{const}$ scaling law. Thus, cryogenic counterparts of these devices are able to operate at frequencies approaching $1$~GHz without excess loss due to phonon-phonon interaction\cite{quartzPRL}, while room temperature operation limits these devices to tens of Megahertz. Secondly, it is observed experimentally and explained theoretically that at cryogenic temperatures {due to significantly higher speed of sound,} longitudinal phonons exhibit lower losses than shear. The situation is the reverse at room temperature, where the quality factor of a typical shear mode is greater than that of a longitudinal one. For example, at cryogenic temperatures, only the A$_{n,0,0}$ modes exhibit quality factors in excess of $10^9$ and $Q\times f$ products on the order of $10^{18}$~Hz, for values of $n$ as high as 227\cite{Goryachev1,ScRep,quartzPRL}.
Moreover, typical quartz plate resonators for room temperature applications are designed to work at a mode that has a frequency-temperature turnover point at slightly elevated temperatures for stress-insensitive crystal cuts. This is essential as temperature and stress insensitivity as well as high-$Q$ are required to achieve very long time frequency stability for the use of acoustic resonators in frequency control applications. This situation long side with high quality factors (over $10^{6}$) is typically achieved with a slow shear 3rd or 5th overtone mode (C$_{3,0,0}$ or C$_{5,0,0}$). At the same time, for cryogenic applications, long term frequency stability is not always required.}
 Thus, for the reasons described above, cryogenic devices are compressional vibration devices whereas room-temperature resonators are transverse mode resonators. 

In order to reduce losses due to phonon tunnelling to the environment, phonon-trapping techniques are used. The two most popular approaches are electrode loading (by optimising ratio between an electrode and the resonator) of the disc centre and the curvature of the plate surfaces. The second technique involves the separation of the electrode from the vibrating plate so that it is non-contacting (BVA resonator)\cite{1537081,besson2}, which eliminates the losses due to electrode loading. 
 The acoustic wave is said to be well trapped if most of its energy is separated from the plate borders, so that its mechanical coupling to the support is minimised. Normally, the resonators are designed to confine the acoustic energy in its centre. Nevertheless, modes with nonzero $m$ and $p$ are possible, although usually the quality factor decreases with increasing $m$ and $p$ due to the higher order mode shapes exhibiting significant coupling to the support at the plate border.

The study of vibrating plates is a long standing subject of research, both experimentally and theoretically\cite{mindlin,szilard,reddy}. In particular, our interests is in the application of high acoustic $Q$ piezoelectric plates\cite{Tiersten:1995fh}. Many authors have contributed to the field optimising the structure, material quality\cite{Brice:1985xd}, crystal cuts\cite{Kusters:2014mn,pz:1988zr}, ageing\cite{Kosinski:1992im}, electrode design\cite{lewis,EerNisse:1975jb}, nonlinearities\cite{nonlin,Goryachev:2zn}, noise properties\cite{Gagnepain:1976vs,Groslambert:1999dg,Goryachev:2012jx}, thermal\cite{Ballato:1978xi,EerNisse:1980mq} and vibrational\cite{Filler:1988oa,Driscoll:2031xz} stabilities etc. Also, a substantial amount of work is dedicated to the analysis of contoured resonators with the ability to trap acoustic energy inside the plate\cite{Holland:1969le,Tiersten:1979sh,Sinha:2001wb}. One of the most influential achievements of this field is known as
Stevens-Tiersten theory\cite{Tiersten:1976hz,Tiers1, Shi:2014ac}. This theory gives a partial differential equation for {the dominant component of the displacement} $u_d$, for the piezoelectric {spherically contoured} BAW cavity{, which has slowly varying thickness in the $x$-$y$ plane due to the large radius of curvature} (Fig.~\ref{F0041FS})\cite{Tiers1}. 
 \begin{equation}
\left. \begin{array}{ll}
\displaystyle \rho\ddot u_d +\frac{\pi^2 n^2 \hat{c}_{z}}{4h_0^2}\Big(1+\frac{x^2+y^2}{2Rh_0}\Big)u_d=\\
\displaystyle =M_n{\partial^2_{xx} u_d}+P_n\partial^2_{yy} u_d + (-1)^{(n-1)/2}\frac{e_{(z)}}{c_{z}}\frac{4\ddot v}{n^2 \pi^2},
\end{array} \right. 
 \label{P117PP}
\end{equation}
Here $n$ is the overtone number, $M_n$ and $P_n$ are parameters, which depending on material constants, $v$ is  the voltage applied across the plate surfaces, $R$ is the resonator plate radius of curvature, $2h_0\ll R$ is the resonator thickness, $\rho$ is the material mass density, {$e_{{(z)}}$ is the effective piezoelectric coefficient, and $c_{z}$ and $\hat{c}_{z}$ are the modified effective elastic coefficients for the longitudinal mode in the crystal of certain crystal orientation, which are different due to the piezoelectric effect. The values are given by
 \begin{equation}
\left. \begin{array}{ll}
\displaystyle c_{z} = \overline{c}_{z}-{e_{(z)}^2}\varepsilon_{z},\hspace{5pt}
\displaystyle \hat{c}_{z} = \overline{c}_{z}-\frac{8}{n^2\pi^2}{e_{(z)}^2}\varepsilon_{z},
\end{array} \right. 
 \label{P117PPx}
\end{equation}
where $\varepsilon_{z}$ is the dielectric constant along $z$, and $\overline{c}_{z}$ is unperturbed effective elastic coefficient when no piezoelectric interaction is present. Such correction terms are typically small, for example, for higher order OTs the correction term for $\hat{c}_{z}$ is negligible\cite{eer}.}
{The dominant component of the displacement $u_d$ is either along $x$, $y$ or $z$ axes (correspondingly $u_x$, $u_y$ or $u_z$) depending on the type of the thickness mode: slow shear, fast shear or longitudinal\cite{eer}. Due to the higher sound velocity, the latter can be excited to much higher OTs and exhibits extremely high quality-factors. In this case only $u_z$ is considered further with the $z$ index dropped.} Also, we drop the external driving term and consider only the internal cavity wave distribution. It should be noted that for simplicity only the case of a square plate is considered further.  

Parameters $M_n$ and $P_n$ have the following dependence on the overtone number:
  \begin{equation}
\left. \begin{array}{ll}
\displaystyle M_n = M + \frac{a_x}{n}\cot{\frac{\kappa_x n\pi}{2}} + \frac{a_y}{n}\cot{\frac{\kappa_y n\pi}{2}},
\end{array} \right. 
 \label{P117PPd}
\end{equation}
where $M$, $a_x$, $a_y$, $\kappa_x$ and $\kappa_y$ are material specific parameters\cite{Tiers1,eer}. {The same type of functional dependence could be written for $P_n$, which is slightly different only in values of the material specific parameters due to the material anisotropy.} It should be noted that both $\kappa_x$ and $\kappa_y$ approach unity in the limit of an isotropic material, since both parameters are defined as square roots of ratios between sound velocities in the different directions\cite{Tiers1}. Thus, for slightly anisotropic devices, $\cot{\frac{\kappa_{i=x,y} n\pi}{2}}\rightarrow 0$ and the last two terms in the expression are negligible.
Since only odd (piezoelectrically excited) overtones in the limit of large $n$ are of the interest for this work, dependence of $M_n$ and $P_n$ parameters on the OT number could be neglected. Thus, in the following sections the corresponding indices will be dropped.

\section{Phonon-Trapping in a Curved Plate}
\label{trap}

Utilising a quasi-particle understanding of bulk acoustic vibration, phonon dynamics in the $x-y$ plane can be considered independent from the resonant behaviour along the $z$-axis. In this picture, the nontrivial geometry of the plate creates a potential well for phonons in the plane of the plate. This potential serves well as a phonon trap, which does not allow phonon escape through the clamping points. This sections present the main characteristics of such phonon traps using classical results based on solutions of equation (\ref{P117PP}).

%Equation (\ref{P117PP}) is an inhomogeneous PDE.
 Implying harmonic motion ${u}(x,y,z,t) = {u}(x,y)\sin\frac{n\pi z}{2h_0}e^{i\omega_{nmp}t}$, the eigensolutions of the homogenous problem corresponding to eq.~(\ref{P117PP}) can be approximated by
 \begin{equation}
 \left. \begin{array}{ll}
\displaystyle u_{nmp} = e^{-\alpha n\pi\frac{x^2}{2}}H_m\big(\sqrt{\alpha n\pi}x\big)
 \displaystyle \times e^{-\beta n\pi\frac{y^2}{2}}H_p\big(\sqrt{\beta n\pi}y\big),
\end{array} \right. 
 \label{P118PP}
\end{equation}
where $H_x$ is a Hermit polynomial and
 \begin{equation}
 \alpha^2 = \frac{\hat{c}_{z}}{8Rh_0^3M},\hspace{10pt} \beta^2 = \frac{\hat{c}_{z}}{8Rh_0^3P}.
 \label{P119PdP}
\end{equation}

\begin{figure}[ht!]
     \begin{center}
            \includegraphics[width=0.45\textwidth]{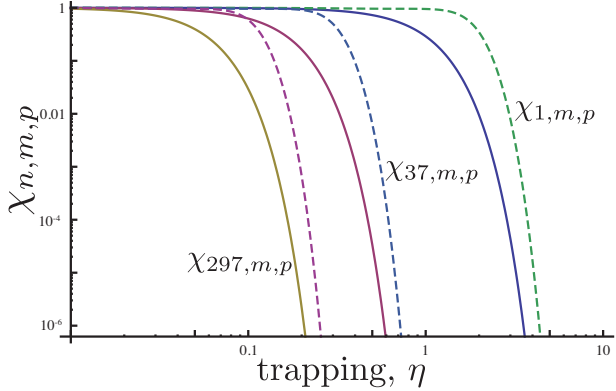}
            \end{center}
    \caption{Phonon tunnelling probabilities $\chi_{n,0,0}$ (solid curves) and $\chi_{n,2,2}$ (dashed curves) for various OTs as a function of trapping parameter $\eta=\eta_x=\eta_y$.}%
   \label{F0041FSd}
\end{figure}

For such a phonon trap, the escape probability could be defined by the amount of vibrational energy in a finite curved plate compared to total energy in a corresponding infinite plate:
 \begin{equation}
\displaystyle \chi_{nmp}^{-1} = 1-\int\limits_{\mathcal{A}} u_{nmp}^2 ds /  \int\limits_{\mathbb{R}^2} u_{nmp}^2 ds
 \label{R002R}
\end{equation}
where $\mathcal{A}\in\mathbb{R}^2$ denotes the area of the finite curved plate in the $x-y$ plane. This definition is based on the analogy between solution (\ref{P118PP}) and the quasi-particle wave-function moving in {a finite harmonic potential well} in the $x-y$ plane. In this description the equation of motion~(\ref{P117PP}) could be transformed to an analogy of the Schr\"{o}dinger equation. Thus, the vibration outside the resonator plate indicates the tunnelling into the environment\cite{Aspelmeyer}, {in an analogous way as parts of a particle wave-function outside a potential well of finite hight would contribute to quantum mechanical tunnelling\cite{tunnelling}.} {Although it should be noted that $\chi_{nmp}^{-1}$ is an approximation for an upper bound of the clamping losses $Q_{\mbox{clamp}}^{-1}$, that neglects the details of how the structure is supported.}
For the two possible combinations $m$ and $p$, the rate is 
 \begin{equation}
 \left. \begin{array}{ll}
\displaystyle \chi_{n00}^{-1} = 1-\mbox{Erf}(\sqrt{n}\eta_x)\mbox{Erf}(\sqrt{n}\eta_y),\\
\displaystyle \chi_{n22}^{-1} = 1-\Big[\mbox{Erf}(\sqrt{n}\eta_x)-\frac{\sqrt{n}\eta_x}{\sqrt{\pi}}(1+2\eta_x^2 n)e^{-n\eta_x^2}\Big]\\
\displaystyle\hspace{32pt}\times\Big[\mbox{Erf}(\sqrt{n}\eta_y)-\frac{\sqrt{n}\eta_y}{\sqrt{\pi}}(1+2\eta_y^2 n)e^{-n\eta_y^2}\Big],
\end{array} \right. 
 \label{R002Rd}
\end{equation}
where $\eta_x = \sqrt{{\pi}\alpha}L$, $\eta_y = \sqrt{{\pi}\beta}L$ are unitless trapping parameters along $x$ and $y$ axis. These parameters describe how well the Gaussian distribution of vibration in $x-y$ fits {within the resonator of width $2L$.} The tunnelling probabilities for $(n,0,0)$ and $(n,2,2)$ types of modes as a function of trapping parameter are shown in Fig.~\ref{F0041FSd}. {Note that since $\eta\sim  L/(R h_0^3)^{1/4}$, trapping is possible when $L\ll R$. }

%\begin{figure}[ht!]
%     \begin{center}
%            \includegraphics[width=0.45\mboxwidth]{escape2.eps}
%            \end{center}
%    \caption{Phonon tunnelling rate $\chi_{n,2,2}$ for various OTs as a function of trapping parameter $\eta=\eta_x=\eta_y$.}%
%   \label{F0042FS}
%\end{figure}

Fig.~\ref{F0041FS} demonstrates that the phonon tunnelling probability drops very fast at a specific value of $\eta$. This value of the trapping parameter can be understood as a trapping threshold. Moreover, at higher OTs this threshold is achieved for lower values of $\eta$. Nonzero in-plane wave numbers $m$ and $p$ always increase the tunnelling threshold. Thus, it is preferable to work with the fundamental OT resonance family ($m = p = 0$) due to its lower coupling to the environment. Increasing both $n$ and $\eta$ increases the focusing of the acoustical vibration to the centre of the plate, which simultaneously reduces the amount of the material in the body involved in the action.
% minima at lower values of $\eta$. 

\section{Cavity Zero-Point Fluctuations}
\label{fluc}

The variance of zero point fluctuations (ZPF) of displacement and momentum of a Harmonic Oscillator are 
\begin{equation}
\big<\hat{x}^2\big> = \frac{\hbar}{2\omega m_{\mbox{eff}}}, \hspace{3pt}\big<\hat{p}^2\big> = \frac{{\hbar\omega m_{\mbox{eff}}}}{2}.%\Bra{0}(a+a^\dagger)^2\Ket{0},
 \label{P007DD}
\end{equation}
Calculations of these values for each mode of the acoustic wave device require knowledge of the mode effective mass and angular frequency. For the curved resonator the parameters depend on the geometry of the surfaces and the mode numbers.

The angular frequencies of thickness modes of a curved plate is approximated as follows:
 \begin{equation}
\omega_{nmp}^2 \approx \frac{n^2\pi^2\hat{c}_{z}}{4h_0^2\rho}\Big[1+\frac{\chi_x}{n}(2m+1)+\frac{\chi_y}{n}(2p+1)\Big]
 \label{P119PP}
\end{equation}
where $\hat{c}_{z}$ is an effective elastic constant for the given type of vibration, $\chi_x = \frac{1}{\pi}\sqrt\frac{2h_0M}{L\hat{c}_{z}}$ and $\chi_y = \frac{1}{\pi}\sqrt\frac{2h_0P}{L\hat{c}_{z}}$. For high-$Q$ BAW cavities the expression can be approximated by just the multiplier term before the square brackets, because in the limit of large $n$, $R\gg2h_0$  and low $m$ and $p$ numbers (usually both are zero) the last two terms in the expression are much less than 1.

The effective mass of an acoustic mode is defined as the sum of masses of all elementary parts $dv$ of the vibrating body scaled by the involvement of these parts into the vibration:
\begin{equation}
m_{\mbox{eff}} = \int\limits_\mathcal{V} \rho \frac{u_{nmp}^2(x,y)}{u^2_{\mbox{max}}}dv,% =\rho h_0\frac{\mbox{Erf}(\sqrt{\alpha n \pi} L)\mbox{Erf}(\sqrt{\beta n \pi} L)}{\sqrt{\alpha\beta} n}
 \label{P016DD}
\end{equation}
where $\mathcal{V}$ is the whole device volume. 
For the case of the main modes ($m=0$, $p=0$), the effective mass is given by the expression:
\begin{equation}
m_{n,0,0}  =\rho {\pi}{h_0L^2}\frac{\mbox{Erf}(\sqrt{n } \eta_x)\mbox{Erf}(\sqrt{n} \eta_y)}{\eta_x\eta_y n}=\frac{\overline{m}}{\xi_n}
 \label{P016bDD}
\end{equation}
where $\overline{m}=4\rho h_0L^2$ is a mass for a corresponding flat plate.
Thus, the effective mass for an acoustic mode within the curved geometry can be represented by an effective mass of the equivalent flat plate scaled by a geometrical factor $\xi$:
\begin{equation}
\xi_n = \frac{4}{\pi}\frac{\eta_x\eta_y n}{\mbox{Erf}(\sqrt{n } \eta_x)\mbox{Erf}(\sqrt{n} \eta_y)},
 \label{P119DD}
\end{equation}
which is a function of the OT number and trapping parameter $\eta$. As a result, the curved resonator effective mass depends on the plate geometry, an example of a flat cavity is an FBAR, which was recently cooled to the quantum ground state\cite{fbar}. 

Combining the results (\ref{P119PP}) and (\ref{P016bDD}), the variance of the ZPF in the case of $\alpha=\beta$ can be written as
\begin{equation}
\big<\hat{x}^2\big> =  \frac{\hbar}{\pi^2L^2 \sqrt{\hat{c}_{z}\rho}}\frac{\eta^2}{\mbox{Erf}^2(\sqrt{n}\eta)},
 \label{P117DD}
\end{equation}
which can be further rewritten as the ZPF of a flat device scaled by a geometrical factor arising due to the surface curvature:
\begin{equation}
 \left. \begin{array}{ll}
\displaystyle \big<\hat{x}^2\big> = \big<\hat{x}^2_{\mbox{flat},n}\big> \frac{\eta^2n}{\mbox{Erf}^2(\sqrt{n}\eta)} =  \big<\hat{x}^2_{\mbox{flat},n}\big>  \xi_{n},\\
\displaystyle \big<\hat{p}^2\big> =  \big<\hat{p}^2_{\mbox{flat},n}\big>  \xi^{-1}_{n},
\end{array} \right. 
 \label{P118DD}
\end{equation}
{where $\big<\hat{x}^2_{\mbox{flat},n}\big>$ and $\big<\hat{p}^2_{\mbox{flat},n}\big>$ are the variances for the $n$th OT of the corresponding flat plate. It should be pointed out that the former is inversely proportional to $n$. This dependence is cancelled out by $\xi_{n}$ for large enough value of $\eta$.}

\begin{figure}[ht!]
     \begin{center}
            \includegraphics[width=0.5\textwidth]{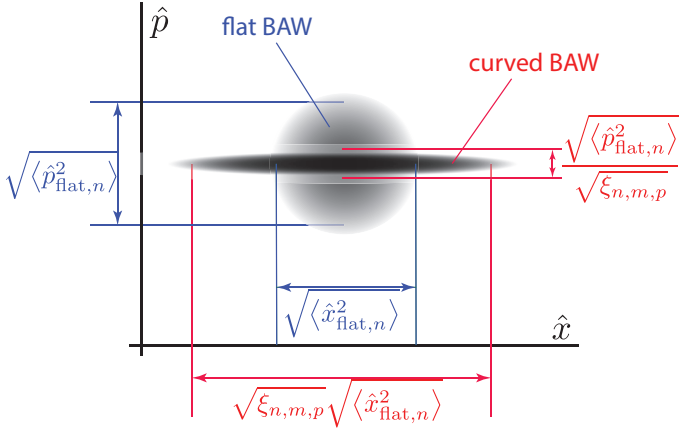}
            \end{center}
    \caption{Influence of the acoustic wave device curvature on the oscillator quadratures. The quadratures are scaled by the square root of the geometrical factor $\xi_{n,m,p}$}%
   \label{F001cFS}
\end{figure}

\begin{figure}[ht!]
     \begin{center}
            \includegraphics[width=0.45\textwidth]{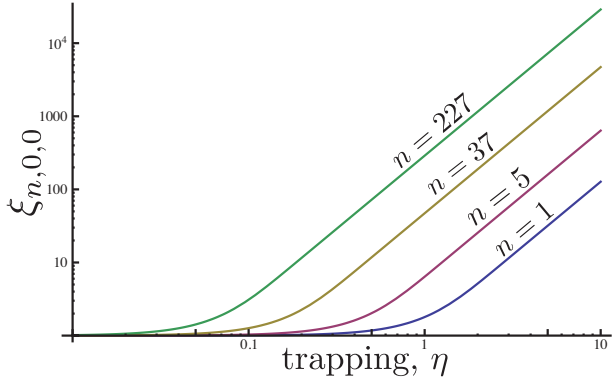}
            \end{center}
    \caption{Dependence of the geometry scale coefficient $\xi$ on the phonon trapping parameter $\eta$ for different OT numbers $n$ and $m=p=0$}%
   \label{F001FS}
\end{figure}

This results demonstrates that the cavity geometry expands one oscillator quadrature and contracts another by changing the effective mass of the mode. This process is shown in Fig.~\ref{F001cFS} where the case of a curved BAW cavity is compared to the corresponding flat device. Dependence of the curvature scaling factor $\xi_{n}$ for $m=0$ and $p=0$ on the trapping parameter $\eta$ and OT number $n$ is shown in Fig.~\ref{F001FS}. The results demonstrate that $\xi_{n,0,0}$ monotonically increases with trapping and the OT number. 

\begin{figure}[ht!]
     \begin{center}
            \includegraphics[width=0.45\textwidth]{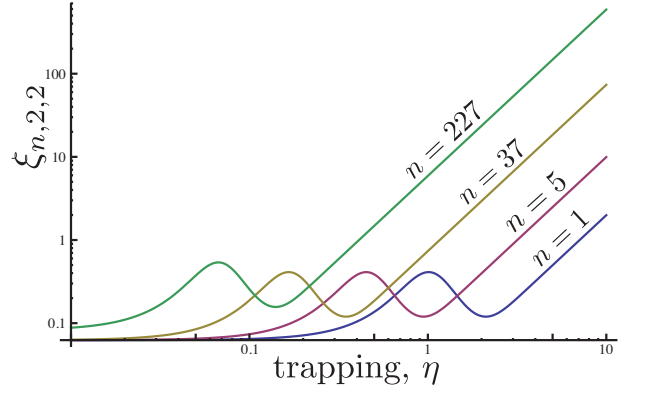}
            \end{center}
    \caption{Dependence of the geometry scale coefficient $\xi$ on the phonon trapping parameter $\eta$ for different OT numbers $n$ and $m=p=2$}%
   \label{F001bFS}
\end{figure}

For non-zero values of $m$ and $p$ and $\alpha=\beta$, the geometrical factor $\xi$ has a few local extrema. For example, for $m=p=2$, the factor is given as follows:
 \begin{equation}
 \left. \begin{array}{ll}
\displaystyle \xi_{n22} =\frac{n\eta^2}{16\pi} \Big({\mbox{Erf}(\sqrt{n}\eta)}-\frac{\sqrt{n}\eta}{\sqrt{\pi}}e^{-n\eta^2}(1+2n\eta^2)\Big)^{-2}.\\
\end{array} \right. 
 \label{R332R}
\end{equation}
This dependence is shown in Fig.~\ref{F001bFS} and exhibits an additional local extremum. These minima correspond to additional nodes of the acoustic wave distribution in the plane. The result suggests that $\xi_{n,0,0}>\xi_{n,2,2}$.

The calculations show that for sufficiently large trapping parameter $\eta$, the variance of zero point fluctuations does not depend on the overtone number $n$. Thus, it is the same for all frequencies. This occurs due to the fact that the mode effective mass is inversely proportional to the overtone number, which cancels the frequency dependence. Moreover, since the effective mass is inversely proportional to the geometrical factor $\xi$, the resulting variance could be geometrically 'amplified' by this factor. 

As a numerical example, we consider a state-of-the-art acoustic cavity, which has been used previously to excite extremely high OTs\cite{quartzPRL}.
For this quartz device $L = 1.5\cdot 10^{-2}$~m, $\hat{c}_{z}\approx105$~GPa (could be varied by changing the cut), $\rho = 2643\frac{\mbox{kg}}{\mbox{m}^3}$, $2h_0=5\cdot 10^{-4}$~m. The material parameter $\hat{c}_{z}$ was calculated from the measured fundamental frequency of the quasi-longitudinal mode $f_{\mbox{fund}}=3.138$~MHz, with a radius of curvature of $R=300$ mm. The resulting displacement and effective mass for the equivalent flat plate is of the following order:
\begin{equation}
 \left. \begin{array}{ll}
\displaystyle\sqrt{\big<\hat{x}^2_{\mbox{flat},1}\big>} \approx 4.7 \cdot 10^{-20}\mbox{m}, \\
\displaystyle\sqrt{\big<\hat{p}^2_{\mbox{flat},1}\big>} \approx   10^{-15}\frac{\mbox{m}\cdot\mbox{kg}}{\mbox{sec}}, \\
\displaystyle \overline{m} = 0.93~\mbox{g},% (\mbox{for }n=1)
\end{array} \right. 
 \label{P120DD}
\end{equation}
which is a subject to scaling by $\sqrt{\xi_{n,m,p}}$, $\sqrt{\xi_{n,m,p}^{-1}}$ and $(\xi_{n,m,p})^{-1}$ respectively, achieved by changing the OT number and the phonon trapping parameter $\eta$ (by changing the radius of curvature). At cryogenic temperatures accessible with modern dilution refrigerators the number of thermal quanta in the fundamental mode at 20 mK is 132, reducing to $0.22$ for the experimentally observed higher order OT of 712.5 MHz with $n=227$ half waves along the thickness of the $z$-axis. 

Numerical calculations of the geometric parameters $\xi$ and $\eta$ requires knowledge of the material dependent parameters $M$ and $P$, which are not exactly known for crystalline quartz at cryogenic temperatures. Although according to the definition\cite{Tiers1}, they are defined as linear combination of various elements of the elastic constant tensor, in particular the ratio between the effective elastic constant $\hat{c}_{z}$ and the transverse elastic parameter $M$. Thus, for order of magnitude calculations it is possible to make an estimation based on the room temperature values of the quartz elastic and piezoelectric tensors\cite{SG}, which gives $\frac{\hat{c}_{z}}{M}\approx0.4$. From this value the trapping parameter $\eta$ is estimated to be approximately $10.7$, which gives a geometrical scaling factor of $10^3$, $5\times 10^3$ and $3.3\times 10^4$ for the $7$th, $37$th and $227$th OTs respectively. In particular, this means that a curved BAW device of these parameters will incorporate modes ranging from gram to microgram mass scales. 

%{The realistic value of the ratio $\frac{c_{z}}{M}$ can be estimated based on room temperature values of quartz parameters giving the value of about $0.4$\cite{SG}. The corresponding value of trapping is $10.7$, and the geometrical scaling factor of the listed above OT are $10^3$, $5\times 10^3$ and $4.4\times10^4$ respectively.}
%Variation of the $\frac{c_{z}}{M}$ factor by plus/minus an order of magnitude gives the geometrical scaling within $3-29$~dB range.

\section{Detection}

Nowadays, the field of precision and quantum measurements is dominated by the optomechanical approach. Due to its extraordinary sensitivity it has been used in various sorts of applications, such as gravitational wave detection and the detection of vibration of mechanical oscillators cooled to near the quantum ground state. However, with the resent results on acoustic wave devices\cite{fbar}, it has been realised that the piezoelectric properties of a material could also be utilised for the conversion of mechanical motion into detectable electrical signals. In this section, we compare these two approaches when applied to the curved BAW device. 

\subsection{Optomechanical Detection}

The optical approach to detect vibrations of a BAW resonator is similar to the detection of the mechanical motion of any other optomechanical systems\cite{Kippen}. The method utilises the motion of part of the system to displace one of the mirrors or a boundary condition for an optical or microwave cavity. In the case of a BAW device, if one of the resonator surfaces is coated with a perfect mirror, the motion of the ideally thin mirror corresponds to the vibration of the plate at the point of crystal-vacuum interface ($u(x,y,2h_0)$ and $u(x,y,0)$). Thus, for odd OTs the detectable displacement corresponds to the values of $u$ calculated in the previous section, since the interface point corresponds to a maximum of acoustic vibration. {As a result for an ideally narrow light beam pointing at the BAW cavity centre, the problem is reduced to the detection of $\sqrt{\big<\hat{x}^2\big>}=\sqrt{\big<\hat{x}^2_{\mbox{flat},n}\big>\xi_n}$, which is enhanced with the increase of the OT number as $\sqrt{n}$ due to geometrical scaling as described in the previous section.} Thus, for this type of a system it is advantageous to work with higher overtones at higher frequencies for this optomechanical type of detection {for two reason: 1) The improvement of the thermal occupancy without a sacrifice of the amplitude of fluctuations: 2) The scaling of the detectable vibration as $\sqrt{n}$. The first reason} contradicts the conventional state of affairs where the magnitude of the zero point fluctuations is usually inversely proportional to the frequency, contrary to the requirement of high frequency for the minimisation of thermal quanta for Bose-Einstein statistics. So, this is overcome with the implementation of the curved BAW devices where both conditions require higher frequencies. 

\subsection{Piezoelectrical Detection}
\label{piezo}

Unlike a typical optomechanical setup, the electro-mechanical coupling involving the piezoelectric effect\cite{pz:1988zr} is {\it not} parametric detection. Piezoelectricity is a linear phenomenon that relates mechanical and electrical field components\cite{Yang:2006pr}. Thus, it is important that by itself it is not sufficient for measurements in the quantum regime. It has to be complemented by some external nonlinear devices such as Josephson Junctions\cite{fbar}.  

It is usually considered that one-port BAW devices are excited by applying time-varying voltage to the electrodes, whereas readout is performed by detecting the resulting alternating current through the device. This current is found as integral of the time-derivative of the z-component of the displacement vector $\mathbf{D}$ over the electrodes $\mathcal{A}_e\in \mathcal{A}$ evaluated at one of the electrodes (e.g. $z=2h_0$):
\begin{equation}
I=-\int\limits_{\mathcal{A}_e}\partial_tD_zds = -e_{\mbox{eff}}\int\limits_{\mathcal{A}_e}\partial^2_{tz}u(x,y)ds
 \label{R120DD}
\end{equation}
where $e_{\mbox{eff}}$ is an effective material constant involving piezoelectric and elastic tensors. This current excludes components due to parasitic capacitance of the electrodes. Taking into account space dependencies of the displacement (\ref{P118PP}), the resulting current is 
\begin{equation}
I_{n,m,p}= e_{\mbox{eff}}\frac{\pi\mu_{n,m,p}}{\sqrt{\alpha\beta}h_0}\partial_t u,
 \label{R121DD}
\end{equation}
where $\mu_{n,m,p}$ is an electrode overlap factor. This parameter tell how much of the acoustic vibration is covered by the electrodes. For the $(n,0,0)$ type of mode, this factor is
\begin{equation}
\mu_{n,0,0} = {\mbox{Erf}\frac{\sqrt{n}\nu_x}{\sqrt{2}}{\mbox{Erf}\frac{\sqrt{n}\nu_y}{\sqrt{2}}}},%{{n}\nu_y\nu_x},
 \label{R122DD}
\end{equation}
where $\nu_x = \sqrt{\pi\alpha}\widetilde{L}$, $\nu_y = \sqrt{\pi\beta}\widetilde{L}$, and $\widetilde{L}<L$ is a characteristic dimension of an electrode {as shown in Fig.~\ref{F0041FS}}. Function $\mu_{n,0,0}(\nu)$ approaches unity when electrodes cover most of the acoustic energy. This dependence is very similar to the inverse of the dependence $\xi_n(\eta)$ although it involves another resonator parameter, electrode dimension $\widetilde{L}$, rather than the plate dimension $L$. In realistic systems, these electrodes are always much smaller than plates themselves. %, although whereas parameter $\eta$ should be maximised, parameter $\nu$ could be greatly reduced by optimising the electrode area. 

The variance of the output current due to the mechanical zero point fluctuations could be found as follows:
\begin{equation}
\sqrt{\big<\hat{I}^2_{n,m,p}\big>}= e_{\mbox{eff}}\frac{\pi\mu_{n,m,p}}{\sqrt{\alpha\beta}h_0\overline{m}}{\sqrt{\xi_{n,m,p}}} \sqrt{\big<\hat{p}^2_{\mbox{flat},n}\big>}.
 \label{R123DD}
\end{equation}
This result demonstrates that despite the fact that $\big<\hat{p}^2\big>$ scales down with the $\xi_{n,m,p}$, the detected current is still proportional to $\sqrt{\xi_{n,m,p}}\sim\sqrt{n}$. This result means that like in the case of optomechanics, piezoelectrical detection also benefits from increase of geometrical factor $\sqrt{\xi_{n,m,p}}$. %Moreover, equation (\ref{R123DD}) suggests that the signal grows with the OT number even faster. 
Thus, like in the case of optical detection, high frequencies are desirable.

It has to be mentioned that operation of BAW devices at extremely high OTs requires the redesign of the electrodes. This is because at higher frequencies the electrodes appear as a parasitic capacitive shunt, which significantly reduces the electro-mechanical coupling. This problem can be overcome by reducing the electrode area in a way that it also has no effect on the overlapping function $\mu_{n,m,p}$ since for higher OTs the vibration becomes more focused at the centre (smaller spot size). Although, in each frequency range there will be a different electrode size that is optimal in terms of coupling. If one defines an optimal value of overlapping $\mu_{\mbox{opt}}$, e.g. three standard deviations ($3\sigma$), it is possible to calculate the optimal electrode size or minimal characteristic dimension $\widetilde{L}$ that reaches the optimal overlapping:
\begin{equation}
\widetilde{L}_{\mbox{opt}} = \frac{L}{\eta}\sqrt{\frac{2}{n}}\mbox{Erf}^{-1}(\sqrt{\mu_{\mbox{opt}}}),
 \label{R123DDe}
\end{equation}
assuming $\alpha=\beta$. The corresponding shunt capacitance could be approximated as a capacitance of two parallel plates formed by the electrodes\cite{lewis}. Thus, the corresponding parasitic shunt capacitance is scaled as $C_0\sim L_{\mbox{opt}}^2\sim (n\eta^2)^{-1}$. As a result the corresponding optimal parasitic shunt impedance is 
\begin{equation}
Z_{\mbox{shunt}}=\frac{1}{i\omega_{n,m,p}C_0(n)}= -i \frac{2h_0^2}{\varepsilon_{z}L^2}\sqrt{\frac{\rho}{\hat{c}_{z}}}\eta^2n^0\mbox{Erf}^{2}(\sqrt{\mu_{\mbox{opt}}}),
 \label{R123DDg}
\end{equation}
where $\omega_{n,m,p}$ is approximated by the first term in (\ref{P119PP}). This result shows that the optimal shunt impedance does {\it not} depend on the OT number but is proportional to the square of the phonon trapping parameter. The latter has to be maximised in order to minimise photon leakage into the environment as explain in Section~\ref{trap}. Thus, coupling to the mechanical mode could be maintained constant for all OTs by optimising the electrode size. 

Taking the numerical example given at the end of Section \ref{fluc} with the estimated value of trapping $\eta=10.7$, the optimal parasitic capacitance for $3\sigma$ coverage is calculated to be $0.5\times n^{-1}$~pF, corresponding to a parasitic impedance of approximately $312$~kOhm. This value should be compared to that of the active impedance of an $RLC$ or Butterworth-Van Dyke model\cite{BVD} for each OT. For SC-cut quartz resonators at cryogenic temperatures, this resistance can be as low as few Ohms at 3-11th overtones and never exceeds $100$ Ohms at higher OTs \cite{mybook,SunFr}. Thus, the optimal parasitic impedance is negligible.

%where the time derivative has been used to go from displacement measurements $\big<\hat{x}^2\big>$ to momentum measurements $\big<\hat{p}^2\big>$. 

%Result (\ref{R123DD}) stats that current measurements of a piezoelectric BAW device are equivalent to momentum measurements. The measurements explicitly incorporate geometrical squeezing factor $\sqrt{\xi_{n,m,p}}$.

\section{Comparison with a Membrane}
 
 It is instructive to compare a curved BAW cavity with a mechanical resonator that is a traditional choice for various experiments involving cooling to the quantum ground state\cite{Thompson:2008ve}. So, we consider motion of a rectangular membrane\cite{Yu:2013ys} ($a$ by $b$) thickness $h\ll \{a, b\}$ made of material of mass density $\rho$ with the stress $\tau$.
The problem of membrane motion is solved by the function:
\begin{equation}
u = A\sin(\alpha x)\sin(\beta y) \exp(-i\omega t),
 \label{P003DD}
\end{equation}
where $\omega^2 = c^2(\alpha^2+\beta^2)$ and from boundary conditions:
%\begin{equation}
$\alpha = \frac{\pi m}{a}, \hspace{5pt} \beta = \frac{\pi n}{b}$, $m,n\in \mathbb{Z}$
% \label{P004DD}
%\end{equation}
giving the angular frequency of mechanical vibration:
\begin{equation}
\omega =\pi c \sqrt{\frac{m^2}{a^2}+\frac{n^2}{b^2}},
 \label{P005DD}
\end{equation}
where $c^2=\tau/\rho$ is effective sound velocity.

The resonator effective mass is calculated as a sum over its surface of masses of its constitutive parts scaled by the involvement into the mode vibration: $m_{\mbox{eff}} = \rho h \frac{ab}{4}$.
%\begin{equation}
%m_{\mbox{eff}} = \int\limits_{\mathcal{A}} \rho h \frac{u^2(x,y)}{u^2_{\mbox{max}}}ds = \rho h \frac{ab}{4},
% \label{P006DD}
%\end{equation}
Unlike the case of the curved BAW resonator, the membrane effective mass is independent of the mode.
Thus, the variance of the zero-point fluctuations is given as follows:
\begin{equation}
\big<\hat{x}^2\big>  = \frac{4\hbar}{\pi \sqrt{\tau \rho} h \sqrt{m^2a^2+n^2b^2}},
 \label{P007DDv}
\end{equation}
where the inverse dependence on the mode numbers $n$ and $m$ is apparent. This means that measurements of higher order modes are unfavourable due to the reduction of the variance of the ZPF when compared to the first order mode  with $n=1$ and $m=1$. As a result the contrary conditions for the choice of mode and frequency arises: the maximisation of (\ref{P007DD}) requires the reduction of frequency, whereas minimisation of the number of thermal quanta requires the inverse. 

%Note that displacement is the only detectable quadrature for the membrane. This quadrate could be detected using optomechanics. 

%Result (\ref{P007DD}) demonstrates that variance of the ZPF is increased proportional to both wave-numbers. Thus, it will be greatly reduced at higher frequencies for higher order modes. 

To compare a membrane and the BAW cavity quantitatively, we choose the membrane of the same size {($a=b=2L$)} made of the material {with the same density with the stress $\tau=105$~GPa}. Such mechanical resonator gives the following results for the variance of the ZPFs, the effective mass and resonance frequency:
\begin{equation}
 \left. \begin{array}{ll}
\displaystyle\sqrt{\big<\hat{x}^2\big>} \approx 6.2 \cdot 10^{-19}\mbox{m}, \\
\displaystyle {m_{\mbox{eff}}} = 7.5\cdot10^{-2}~\mbox{g},\\
\displaystyle f_{\mbox{res}} = 149~\mbox{kHz}
\end{array} \right. 
 \label{P120DDd}
\end{equation}
for the lowest order mode $n=1$, $m=1$. At typical cryogenic temperatures for such experiments ($20$~mK), the average number of thermal quanta is 3230, which requires additional mode cooling resulting in a loss of quality factors due to the associated damping.

%Membrane parameters: $a =b= 10^{-2}$~m, $h = 3\cdot 10^{-7}$, $m=1$, $n=1$, $\tau=5$~GPa (could be specified), $\rho = 3210\frac{\mbox{kg}}{\mbox{m}^3}$ (Silicon Carbide),
%\begin{equation}
%\sqrt{\big<\hat{x}^2\big>} = \Big(\frac{4\hbar}{\pi \sqrt{2\tau\rho} h a}\Big)^{0.5} = 1.5\cdot 10^{-16} \mbox{m}.
% \label{P008DD}
%\end{equation}
%for the following frequency and effective mass
%\begin{equation}
%f=\frac{1}{a}\sqrt{\frac{\tau}{2\rho}} = 88.3 \mbox{kHz, } m_{\mbox{eff}} = 2.4 \cdot 10^{-5}\mbox{g.}
% \label{P008WD}
%\end{equation}

\section*{Conclusion}

This work demonstrates significant differences of the near ground state behaviour of curved BAW cavities and mechanical resonators. The differences arise due to the focusing of the phonon vibration in the centre of the disk cavity. The focusing is enhanced with the increase of the OT number and correspondingly with the resonance frequency. As a result, the effective mass of vibration drops significantly allowing simultaneous existence of gram and microgram-scale modes in the same macroscopic device. Another consequence of the curvature focusing is the change of the Zero Point Fluctuations of both quadratures of the equivalent Harmonic Oscillator. In particular, it enhances the displacement quadrature resulting in an enhanced magnitude of the Zero Point Fluctuations as the mode frequency increases, allowing easier detection. In addition, BAW devices allow another way of detecting the vibration apart form the traditional optomechanics, such devices are naturally characterised using piezoelectric electro-mechanical conversion. It is demonstrated that the electromechanical conversion of the displacement is also enhanced by trapping, in the same way as in the optomechanical case. Additionally, it is shown that by implementing the optimal electrode size for each specific mode, means that the electromechanical coupling does not depend on the OT number leading to the possibility of optimising the electrodes to work at higher frequencies. This optimal electrode size could be further minimised by increasing the trapping parameter. 

\section*{Acknowledgements}
This work was supported by the Australian Research Council Grant No. CE11E0082 and FL0992016. Authors are thankful to Serge Galliou, Shlomi Kotler and Pavel Bushev for fruitful discussions. 
\hspace{10pt}

\section*{References}
%\bibliography{biblioBAW}

\providecommand{\newblock}{}

\end{document}